# Fabrication of high quality sub-micron Au gratings over large areas with pulsed laser interference lithography for SPR sensors

Alexander Arriola,[1,2,*] Ainara Rodriguez,[3] Noemi Perez,[1] Txaber Tavera,[1] Michael J. Withford,[2] Alexander Fuerbach[2] and Santiago M. Olaizola[1]

[1]*CEIT and Tecnun, University of Navarra, Pº Manuel de Lardizabal 15, Donostia-San Sebastian, 20018, Spain*
[2]*Centre for Ultrahigh-bandwidth Devices for Optical Systems, Department of Physics and Astronomy, Macquarie University, NSW 2109, Sydney, Australia*
[3]*CIC Microgune, Microsensors Unit, Pº Mikeletegi 48, Donostia-San Sebastian, 20009, Spain*
[*]*aarriola@ceit.es*

**Abstract:** Metallic gratings were fabricated using high energy laser interference lithography with a frequency tripled Nd:YAG nanosecond laser. The grating structures were first recorded in a photosensitive layer and afterwards transferred to an Au film. High quality Au gratings with a period of 770 nm and peak-to-valley heights of 20-60 nm exhibiting plasmonic resonance response were successfully designed, fabricated and characterized.

**1. Introduction**

During the last decades, research on periodic structures on the nanoscale and their applications has increased remarkably. Fields such as nanotechnology and biotechnology have already started to take advantage of the benefits of these structures for biomaterials [1], antireflective optical elements [2] and biosensors [3]. In the latter area, plasmons [4] and plasmonic sensors have been used for instrumentation development and for characterizing biomolecular interactions, respectively. The basic principle of this plasmon based refractive index sensor was introduced by Nemova et al. [5]. The most extensively used configurations for the excitation of plasmons are the Kretschmann configuration [6,7] and the grating configuration [3]. The latter is of particular interest for the fabrication of sensors as it allows a compact architecture and integration. However, in order to get good plasmon resonance and sensor performance, high-quality metallic gratings over large areas are needed [8].

Depending on the quality, period and total surface area of these micro- and nanoscale periodic structures and the materials to be used, a number of different fabrication techniques can be considered, ranging from conventional lithographic techniques to non-conventional methods. Conventional techniques include deep ultraviolet lithography (DUV), electron beam lithography (EBL) and ion beam lithography (IBL). With DUV lithography [9,10] tiny nanoscale features as small as 90 nm can be fabricated but the process is dependent on the availability of masks and is therefore reliant on other lithography techniques such as electron beam lithography. EBL offers the possibility of fabricating small features of about 20nm [11] and gratings [12], but the technique is usually limited to small areas as it is extremely time consuming [13]. A variation of the EBL technique is Ion Beam Lithography (IBL), which uses an ion beam instead of an electron beam. This results in an improved accuracy of the features and complex 3D structures can be realized [14], but the method has the same drawbacks as EBL. In particular, all these techniques are usually limited to areas ranging from a few $\mu m^2$ to a few hundreds of $\mu m^2$.

Non-conventional methods include nanoimprint lithography (NIL) and interference lithography (IL). NIL is a high quality and low cost technique based on the replication of features inscribed in a mould or stamp (often using e-beam) [15-16] that offers simplicity and low cost but the fabrication of the moulds is relatively slow (typical writing speeds of 1–2 $\mu m/s$ [17]). Therefore, this technique is, on the one hand, expensive as moulds must be replaced (after 100 uses [18]) and, on the other hand, poorly suitable to large area processing. IL (or LIL: Laser IL) is a large area non-contact nanofabrication technique that uses interference patterns to structure different materials and that is commonly used to fabricate periodic structures such as gratings [3, 19, 20] and arrays of nanowires [21]. One of its main features is the possibility to change the fabricated period by simply changing the angle of incidence between the interfering beams. The total processed area depends on the beam size and can as such be in the order of up to a few $cm^2$.

LIL based on the classical Lloyd mirror configuration using expanded CW lasers has successfully been demonstrated in the past [22,23]. This approach utilizes a lithographic process on a photoresist to define high-resolution sub-micrometer periodic structures. Nevertheless, these patterns are very sensitive to mechanical vibrations during fabrication as the exposure time of the photoresist can range from a few seconds up to a few minutes [23]. An alternative approach is thus to use pulsed radiation instead of CW lasers. As the exposure time is much shorter (single pulse exposure), problems due to mechanical vibrations can be avoided. Pulsed nanosecond [24,25], picosecond [26] and femtosecond [27] laser sources have all been used to define metallic gratings using the direct-write technique, i.e. by direct ablation without the additional use of a photoresist layer. However, the above mentioned pulsed laser techniques impose limits on the minimum period that can be fabricated in metallic layers as the thermal deposition makes the metallic layer melt and thus destroys any periodic pattern in the surface [28]. As a result, direct-write laser interference patterning is typically limited to periods of a micrometer or larger.

As seen above, while some techniques are useful for fabricating small grating periods over small areas (such as DUV, EBL, IBL) and others for the production of large grating areas with relatively large periods (LIL, direct writing), none of them combines the possibility of fabricating sub-micron metallic gratings over large areas. In this paper we report on the fabrication of high-quality and high-resolution submicron gratings over large areas (over 80 $mm^2$ with less than 0.25% deviation in period over the entire sample) in Au thin films leading to a strong plasmon resonance effect. This was achieved by utilizing a new approach that combines nanosecond pulsed laser interference lithography on a photoresist layer with standard etching techniques allowing a precise control of the depth of the fabricated gold grating. Our novel technique allows to exploit the benefits of previous approaches but at the same time makes it possible to avoid the associated problems. We show that this can only be realized by defining a perfectly balanced set of process parameters. Finally, we demonstrate the application of the gratings for surface plasmon excitation in gold films.

This technique can also be used to fabricate resonant dielectric gratings for evanescent field biosensing applications with sensitivities over 130nm/RIU [29] that can be applied in high throughput devices.

## 2. Fabrication

We have developed a novel fabrication process that is based on high-energy laser interference lithography. Figure 1 shows an overview of the process: First, an Au thin film was deposited on borosilicate glass substrates. Then, via laser interference lithography, the grating pattern was transferred to a negative photoresist spin-coated on top of the metallic layer. A Nd:YAG pulsed nanosecond laser was used as most of the photoresists have the absorption peak at 365 nm and tripling this laser in frequency offers a close enough wavelength (355 nm) to correctly expose the photoresist. The next step was to transfer the pattern from the photoresist to the Au thin film using an etching process. Finally, the remaining resist was removed. The result is an Au thin film with a well-defined grating on the surface that can be used to excite surface plasmons.

The use of pulsed lasers to form an interference pattern and to process a photoresist has technological challenges, as the resist-laser interaction on such small timescales is being explored nowadays [30]. The optimal pulse energy necessary to conveniently pattern a resist depends on the target period and the resist thickness, and must be customized for each process. Pulses with an energy that is too high may burn the resist and/or produce cross-linking of the resist destroying the interference pattern. By comparison, the use of too low energy pulses makes it impossible to process the resist. Therefore, it is necessary to find a compromise where both competing requirements can be met simultaneously. Finally, the fabrication is completed with an etching process in order to transfer the periodic pattern from the photoresist to the underlying Au film.

Using Atomic Force Microscopy (AFM) after every fabrication step, a quantitative value of the period and depth of the grating was obtained. In addition, AFM also provided qualitative information about the amount of photoresist that was still present on the surface as high roughness values and random peaks appear when there is some residual resist.

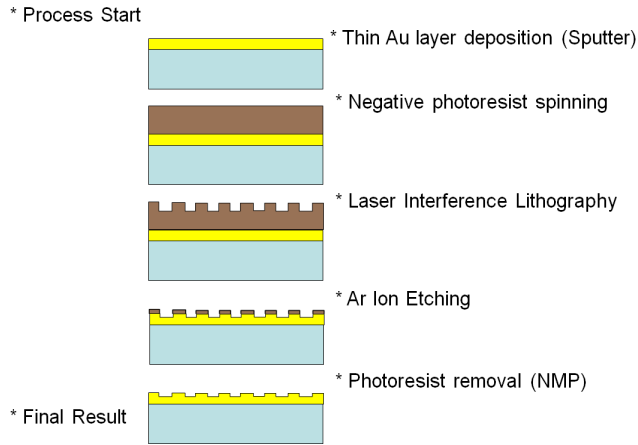

*Figure 1: Step-by-step schematic of the fabrication process*

In the following paragraphs we provide detailed information about each individual fabrication step.

Borosilicate glass substrates (Corning® Eagle 2000™) were first cascade rinsed by fresh acetone and then cleaned with successive acetone and isopropyl alcohol wetted polypropylene swabs. After the cleaning process, a gold film with a thickness of about 60-80 nm (with a measured surface roughness of about 3 nm) was sputtered onto the sample by RF magnetron sputtering (Edwards ESM100 Sputtering System); no adhesion layer has been used between the gold and the glass substrate. Then, a UV-sensitive negative photoresist (AZ®nLOF 2070, MicroChemicals GmbH) diluted with ethyl-lactate at a ratio of 1:3 was spin-coated in order to obtain a photoresist layer target thickness of 300 nm. The diluted photoresist was spun at 4000 rpm for 35 seconds. A pre-exposure bake was performed at 100°C for 1 minute to remove the solvent.

The LIL technique was used for the photoresist exposure. The light source used in the 2-beam interference lithography setup was a frequency tripled Nd:YAG nanosecond laser (355 nm) with a repetition rate of 1 Hz and a beam size of about 1 $cm^2$. As the angle of incidence of the beams determines the period of the resulting structures, it was set to 19.7° with respect to the normal of the plane of the sample in order to obtain a grating period of 770 nm. The photoresist was then exposed to a single pulse of 8 ns with a pulse energy of 15 mJ. After irradiation, a post-exposure bake was performed at 110°C for 1 minute and the sample was developed in $AZ^®$ 726 MIF developer (MicroChemicals GmbH) for 35 seconds at 21°C. Immersing the sample in isopropyl alcohol for 60 seconds stopped the development process. The sample was subsequently immersed in de-ionized water for 60 seconds and finally dried with nitrogen.

Using this method we were able to obtain gratings that showed a very high level of uniformity and periodicity. The measured peak-to-valley depth was 150 nm. However, the shape in the bottom of the structures suggests that the AFM tips could not reach the bottom of these extremely narrow gratings so the depth was probably larger. As can be seen in Figure 2, a shoulder-shaped lift can be observed on the left hand side of each grating period. We believe that this is due to a standing wave that is formed during laser exposure.

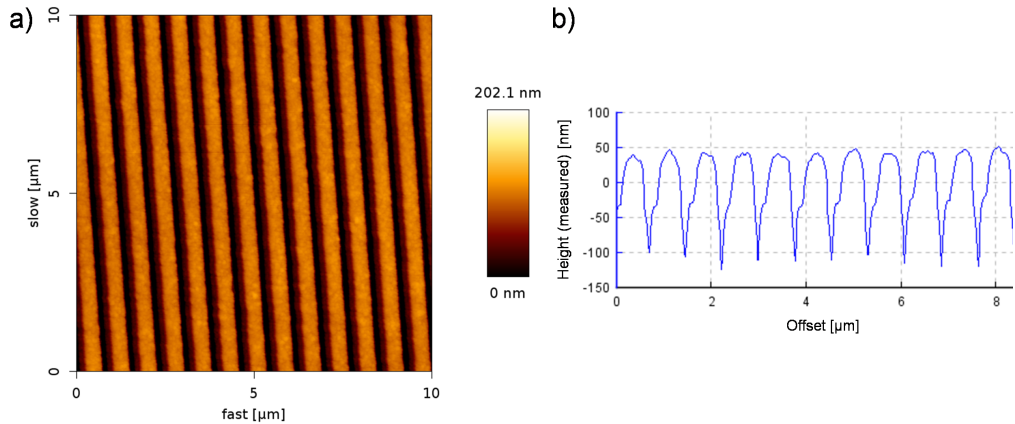

*Figure 2: Topography of the grating inscribed on the spin-coated negative photoresist layer (with a pulse energy of 15 mJ) and its corresponding cross section (b).*

The next step was to replicate the photoresist pattern onto the Au thin film. To transfer the periodic structure, dry etching with Ar ions was used. Power levels lower than the ones specified on the Oxford Instruments' datasheet (300 W) to etch Au were needed in our case because at higher powers (over 200 W) the photoresist reticulated, making it very difficult to remove. Thus, new etching process parameters were established for power levels below 200 W. The total etching time was 390 seconds using the etching parameters of an RF power of 175 W, a pressure of 10 mTorr and an Ar gas flow of 20 standard cubic centimeters per minute (sccm). At this point, a good definition and uniformity of the grating could still be observed and a decrease in grating depth was measured. This is related to a thinning of the photoresist layer as a result of the Ar etching. The shoulder effect described above could no longer be observed after the etching process (see Figure 3).

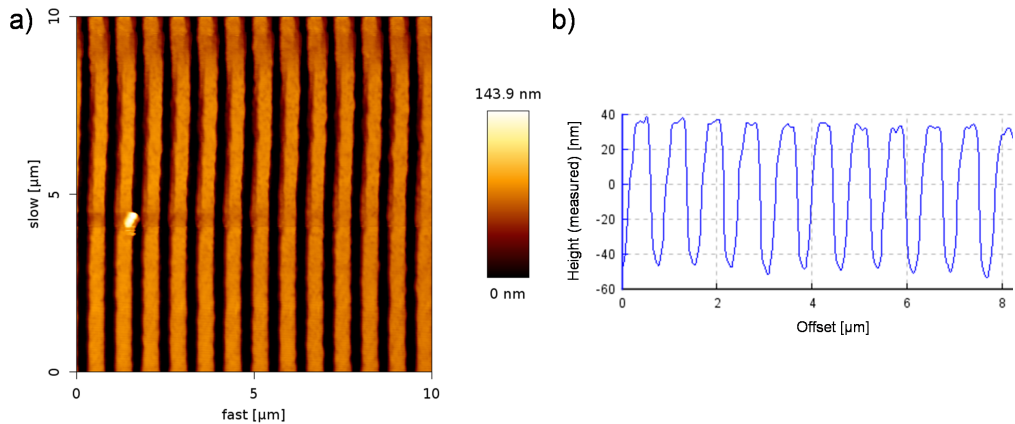

*Figure 3: Topography of the grating after the Ar ion etching process (a) and its corresponding cross section (b).*

The remaining photoresist was removed by an ultrasonic N metil-2-pirrolidone (NMP) bath at 50 °C for 5 minutes, followed by two subsequent 5-minute ultrasonic baths with isopropyl alcohol at 50°C and, with distilled water at 50°C. As can be seen in Figure 4, which shows the topograph of the sample after the NMP cleaning process, a period of 770 nm and a grating depth of about 55-60 nm was obtained. The period of the structures was measured in 5 different areas within the sample (center and all 4 sides) using an optical surface profiler (Wyko® NT 9800). For each area, the period was averaged from the measurement of 30 fringes and then compared to the rest of the calculated values in the other analyzed areas. The results show a variation of less than 0.25% in the grating period.

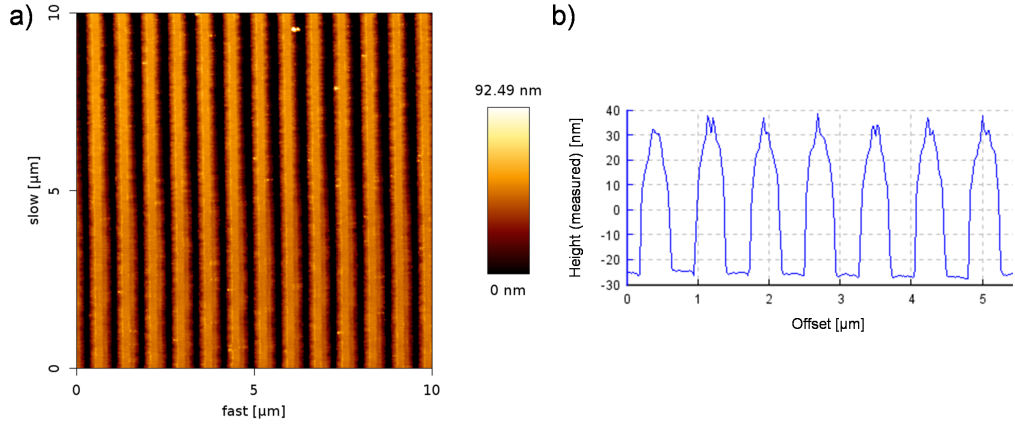

*Figure 4: Topography of the final grating inscribed on the Au surface (a) and its corresponding cross section (b).*

To ensure that there was no photoresist on the sample, an energy dispersive spectroscopy (EDS) analysis was performed, which confirmed that no organic materials were present on the sample.

### 3. Optical properties of the fabricated structures

In order to optically characterize the gold structures, the angle-resolved reflectivity technique was used. This technique allows the characterization of the gratings at different angles of incidence and for different polarizations of incident light. A 50 W halogen lamp was used as a white light source. The light beam passed through a Glan-Thomson polarizer and an iris diaphragm, and was then collimated and directed to the sample. The grating area to be analyzed was selected with an iris diaphragm mounted on a 2-axis translation stage. The light that was reflected off the selected area entered a monochromator and spectrograph (ORIEL MS257) and finally a CCD camera that recorded the signal. In order to vary the angle of incidence of the probe beam, the sample was mounted on a θ/2θ configuration rotational stage.

When illuminating the grating structures with white light at different angles of incidence, a polarization dependence of the reflected signal was observed. The reflectance for s-polarized light showed no noticeable angle-dependant response (see Figure 5) while in the case of p-polarized light a clear angle dependency was observed.

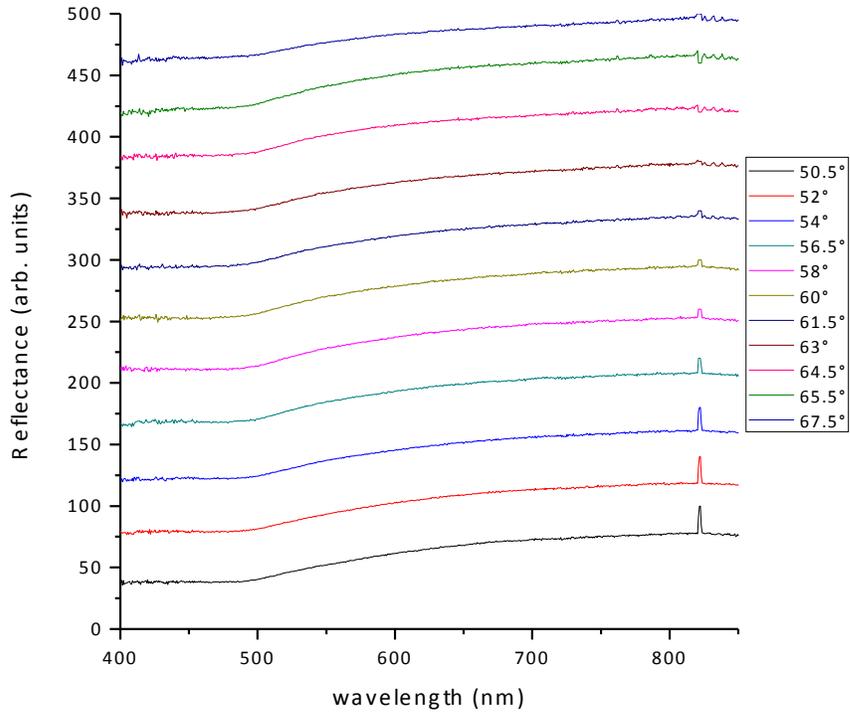

*Figure 5: Reflectance spectra for s-polarized light at different angles of incidence (the individual measurements are offset for clarity).*

As can be seen in Figure 6, pronounced resonance peaks appeared in the reflection spectra that shift towards longer wavelengths for larger angles of incidence.

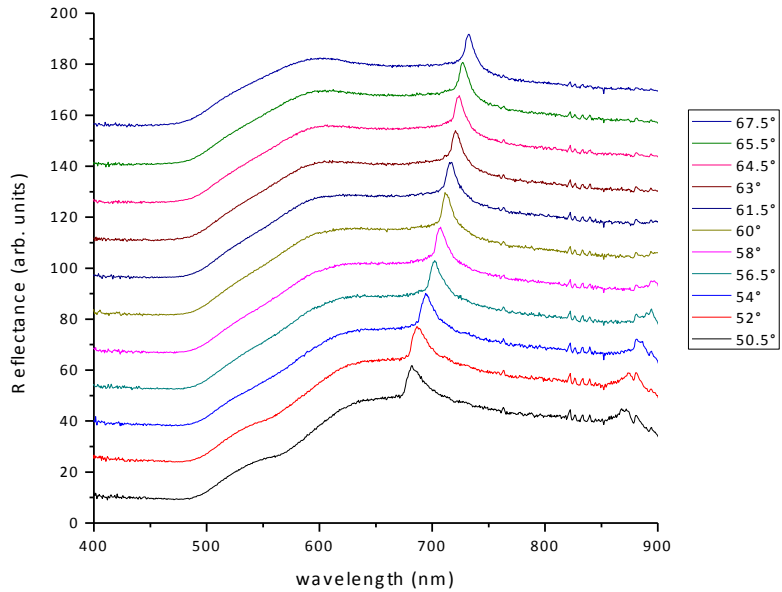

*Figure 6: Reflectance spectra for p-polarized light at different angles of incidence (the individual measurements are offset for clarity).*

The physical mechanism that underpins the observed reflection-enhancement is not entirely clear yet. However, in the last years, several authors have reported surface plasmons having a negative role on transmission when illuminating thin metallic gratings [31-33]. These transmission anomalies seem to be due to a coupling-decoupling effect of the surface plasmons between the upper and lower metal-dielectric interfaces [32], but there is not yet a consensus on the exact physical mechanism causing those transmission anomalies [33]. Notwithstanding, the position of the observed peaks in Figure 6 is determined by the wavelength λ at which surface plasmons are excited in a resonant manner. This wavelength can be calculated using the following equation [34]:

$$\sin\theta = \pm 1 - \frac{m \cdot \lambda}{\Lambda \cdot n_d}$$

where θ is the angle between the surface normal and the incident ray, m is the order of the plasmon resonance (m=-2 in the case of Figure 6), Λ is the period of the gratings and $n_d$ is the refractive index of the dielectric material ($n_d$=1, for air, in this case). When comparing the position in wavelength of the resonance peaks, a perfect agreement between theory and experiments was observed, which reflects the great quality of the grating. In order to show the consistency in the results, a number of samples where analyzed to compare the theoretical with the experimental results. The process parameters for all this samples are inside the narrow window of conditions that allow the successful fabrication of the gratings and are shown in Table 1.

| Sample name | Pulse Energy (mJ) | Etching time (sec.) |
|---|---|---|
| **A1** | 14.26 | 390 |
| **A2** | 18.11 | 390 |
| **A3** | 20.96 | 390 |
| **A4** | 24.64 | 420 |
| **A5** | 31.39 | 420 |

*Table 1: Process parameters for the samples in figure 7.*

The position of the peaks at different wavelengths and for different angles perfectly matches the theory for all investigated orders (+1, -1 and -2), which is due to the high quality and low uncertainty in periodicity of our samples. Figure 7 shows, on the one hand, the theoretical position of the plasmon resonance (represented by the continuous lines) and, on the other hand, the position in wavelength of the resonance peaks for each angle of incidence (represented by symbols).

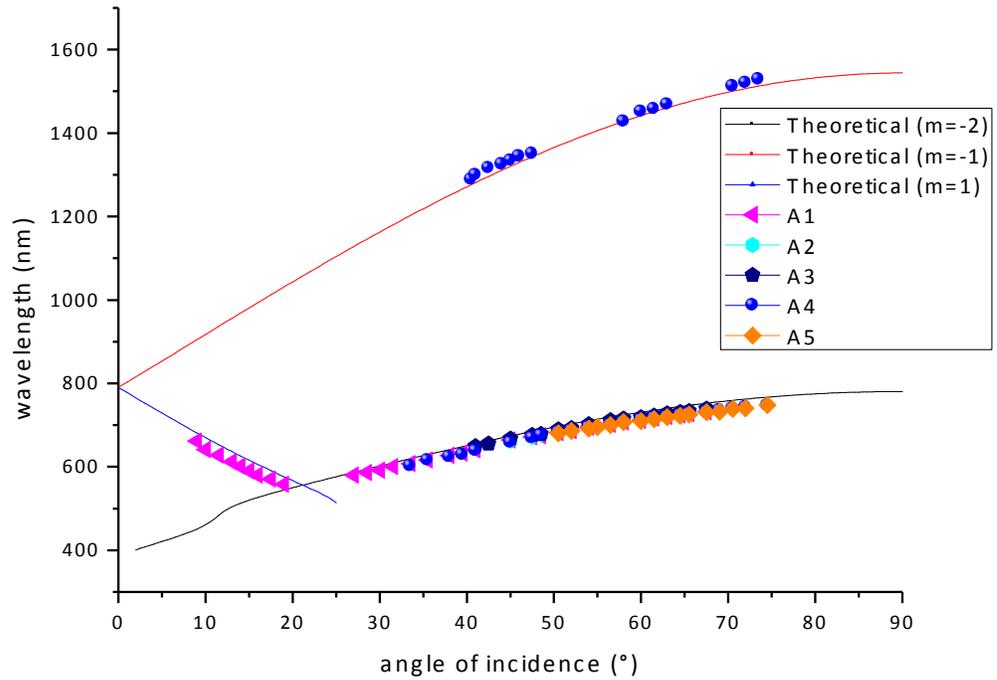

*Figure 7: Comparison between plasmon resonance theory and experiments when the gratings are illuminated with a white light source. Continuous lines show the theoretical position of the resonance peaks while the dots show the experimental position of the peak at different wavelengths*

**4. Conclusions**

In conclusion, we have demonstrated the potential of high-energy pulsed LIL technology for fabricating high quality sub-micron metallic gratings over large areas that exhibit a strong plasmonic response. A fabrication process has been developed and optimized based on first recording the periodic structures in a photosensitive layer using high energy pulsed laser interference lithography followed by transferring the fabricated pattern to a metallic layer. To demonstrate the capabilities of this novel approach, 770 nm period metallic gratings have successfully been fabricated on Au thin films (60-80nm) over large areas of about 80 mm$^2$ with deviations in grating period over the entire sample as low as 0.25%.

Angle resolved reflectance measurements have confirmed the presence of a plasmonic response that perfectly matches the results as predicted by theory.